# Pediatric Appendicitis Detection from Ultrasound Images

Fatemeh Hosseinabadi[1], Seyedhassan Sharifi [2]

[1] *Assistant Professor of Radiology, Zahedan University of medical Sciences, Iran*

[2] *Pediatric Cardiology Subspecialist, Day General Hospital, Iran*

**Abstract**

Pediatric appendicitis remains one of the most common causes of acute abdominal pain in children, and its diagnosis continues to challenge clinicians due to overlapping symptoms and variable imaging quality. This study aims to develop and evaluate a deep learning model based on a pretrained ResNet architecture for automated detection of appendicitis from B-mode ultrasound images. We used the Regensburg Pediatric Appendicitis Dataset, which includes ultrasound scans, laboratory data, and clinical scores from pediatric patients admitted with abdominal pain to Children's Hospital St. Hedwig in Regensburg, Germany (2016–2021). Each subject had 1–15 ultrasound views covering the right lower quadrant, appendix, lymph nodes, and related structures. For the image-based classification task, ResNet was fine-tuned to distinguish appendicitis from non-appendicitis cases. Images were preprocessed by normalization, resizing, and augmentation to enhance generalization. The proposed ResNet model achieved an overall accuracy of 93.44%, precision of 91.53%, and recall of 89.8%, demonstrating strong performance in identifying appendicitis across heterogeneous ultrasound views. The model effectively learned discriminative spatial features, overcoming challenges posed by low contrast, speckle noise, and anatomical variability in pediatric imaging.

## 1. Introduction

Acute appendicitis is the most common surgical emergency in children and adolescents, accounting for approximately 1–2 cases per 1,000 individuals annually. It typically results from luminal obstruction of the appendix, leading to inflammation, bacterial overgrowth, and eventual perforation if left untreated. The condition can progress rapidly from uncomplicated to complicated appendicitis, potentially resulting in peritonitis, abscess formation, and life-threatening sepsis. In pediatric patients, early and accurate diagnosis is essential because younger children often present with atypical symptoms, making clinical differentiation from other causes of abdominal pain more challenging. Studies have reported that diagnostic errors in appendicitis remain a significant concern, with misdiagnosis rates ranging from 15% to 30% in some age groups. False-negative diagnoses increase the risk of complications, while false-positive diagnoses may lead to unnecessary appendectomies, prolonged hospitalization, and higher medical costs. Consequently, improving diagnostic precision is a priority in pediatric emergency medicine [1,2].

Current diagnostic workflows for suspected appendicitis combine clinical evaluation, laboratory testing, and imaging. Clinical scoring systems, such as the Alvarado Score and Pediatric Appendicitis Score (PAS), integrate symptoms (e.g., pain migration, nausea, tenderness) and laboratory markers (e.g., leukocytosis, elevated C-reactive protein) [3]. Although these scores are helpful for risk stratification, they are not definitive and often require confirmation through imaging studies. Among imaging modalities, ultrasound (US) is the preferred first-line technique for children due to its noninvasive nature, lack of radiation, and availability in emergency settings. However, ultrasound diagnosis of appendicitis is highly operator-dependent and subject to variability in patient anatomy, bowel gas interference, and body composition. Visualization of the appendix is successful in only 60–80% of cases, and diagnostic accuracy can drop significantly in obese or uncooperative children. Even experienced radiologists may encounter difficulty differentiating early appendicitis from mesenteric lymphadenitis or gastrointestinal infections, particularly when image quality is suboptimal. These limitations create an urgent need for automated image interpretation systems capable of assisting clinicians with consistent and objective diagnostic insights [4,5].

In recent years, artificial intelligence (AI) and machine learning (ML) have revolutionized diagnostic imaging by allowing computational models to identify complex visual and statistical patterns beyond human perception [6-9]. In radiology, AI has shown significant potential in tasks such as tumor detection in MRI and CT, lung pathology screening in chest X-rays, and cardiac function assessment in echocardiography [10,11]. For ultrasound imaging specifically, AI algorithms have been successfully applied to fetal growth monitoring, thyroid nodule classification, and liver fibrosis staging [12-15]. The main advantage of AI-driven systems lies in their ability to learn directly from raw image data, thereby minimizing dependence on subjective interpretations. Deep learning, particularly Convolutional Neural Networks (CNNs), has emerged as the cornerstone of medical image analysis due to its ability to automatically extract hierarchical features from low-level edges and textures to high-level structural patterns that correspond to anatomical and pathological cues. This makes CNNs particularly suitable for ultrasound data, where signal-to-noise ratios are low, and manual feature engineering is often inadequate.

Among various CNN architectures, Residual Networks (ResNets) have demonstrated outstanding performance in both general computer vision and medical image analysis. ResNets introduce shortcut (skip) connections that bypass one or more layers, allowing the network to learn residual mappings instead of direct transformations. This study aims to harness the power of deep residual learning to improve the diagnostic accuracy of pediatric appendicitis detection using ultrasound data. We employed the Regensburg Pediatric Appendicitis Dataset, a comprehensive dataset collected from pediatric patients admitted with abdominal pain between 2016 and 2021 at the Children's Hospital St. Hedwig in Regensburg, Germany. The dataset includes B-mode ultrasound images, laboratory findings, clinical scores, and expert annotations. Our primary goal was to train a ResNet-based CNN to classify ultrasound images as appendicitis or non-appendicitis and to assess its diagnostic performance against standard metrics.

## 2. Method

### 2.1 Dataset Description

This study utilized the Regensburg Pediatric Appendicitis Dataset, a curated clinical dataset collected retrospectively from pediatric patients admitted with abdominal pain to Children's Hospital St. Hedwig, Regensburg, Germany, between 2016 and 2021. The dataset includes a rich combination of imaging, clinical, and laboratory information to support multimodal diagnostic modeling. Each patient record may contain one to fifteen B-mode ultrasound (US) images, captured from multiple abdominal regions of interest such as the right lower quadrant (RLQ), appendix, intestinal loops, lymph nodes, free fluid areas, and reproductive organs. The ultrasound images are stored in BMP format under the US_Pictures/ directory, with filenames corresponding to subject identifiers and view indices (e.g., 23.7.bmp for patient 23, view 7) [16,17].

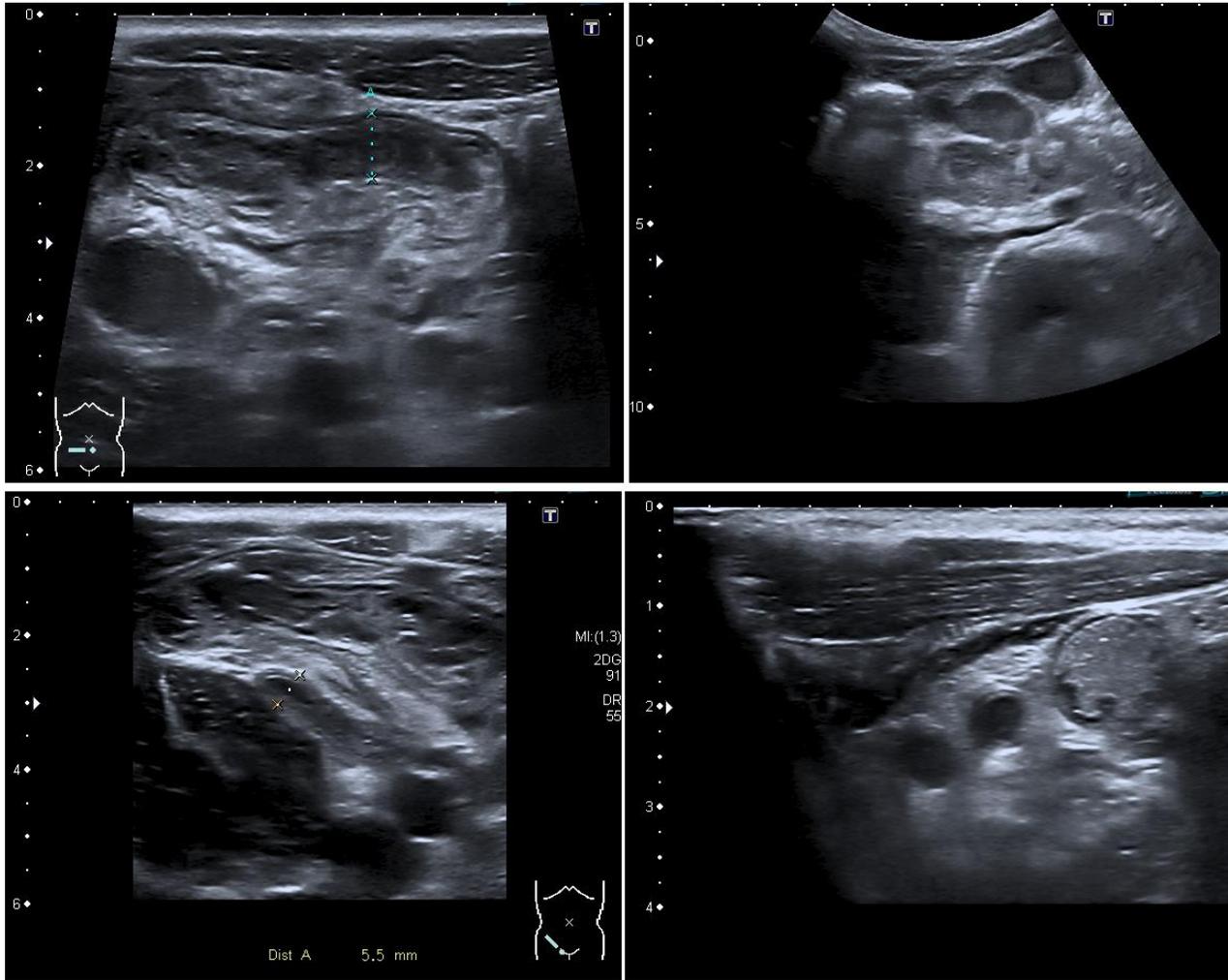

*Figure 1 Representative ultrasound images from the Regensburg Pediatric Appendicitis Dataset. Top left: Ileitis showing bowel wall thickening and inflammation. Top right: Mesenterial lymphadenitis with multiple enlarged lymph nodes in the right lower quadrant. Bottom left: Appendix with surrounding tissue reaction, indicating periappendiceal inflammation and fat echogenicity. Bottom right: Appendix, visualized as a non-compressible tubular structure consistent with acute appendicitis [16,17].*

In addition to imaging data, the accompanying file app_data.xlsx contains tabular variables summarizing laboratory test results, physical examination findings, and expert ultrasonographic assessments. Clinical scoring systems such as the Alvarado Score and Pediatric Appendicitis Score (PAS) were included to contextualize imaging results with established diagnostic criteria. Each subject is annotated for three outcome variables:

- Diagnosis – appendicitis vs. no appendicitis,
- Management – surgical vs. conservative treatment, and
- Severity – complicated vs. uncomplicated (or no appendicitis).

The study was approved by the Ethics Committee of the University of Regensburg (no. 18-1063-101, 18-1063_1-101 and 18-1063_2-101) and was performed following applicable guidelines and regulations. The ethics committee confirmed that there was no need for written informed consent for the retrospective analysis and publication of anonymized routine data according to Art. 27 para. 4 of the Bavarian Hospital Law. For patients followed up after discharge, written informed consent was obtained from parents or legal representatives. In this work, only the Diagnosis label (appendicitis vs. no appendicitis) was used for binary image classification.

## 2.2 Data Preprocessing

The ultrasound images exhibited high inter-patient variability in acquisition parameters such as brightness, contrast, and spatial scale. To standardize inputs, each image was converted to grayscale, resized to 224 × 224 pixels, and normalized to zero mean and unit variance. Data augmentation was applied to improve model generalization and simulate clinical variability, including random rotations (±10°), horizontal flips, contrast adjustments, and Gaussian noise injection. Because multiple images were available per subject, all images were treated as independent samples while ensuring that images from the same patient were confined to either the training or testing split to avoid data leakage. The final dataset was divided into 80% for training and 20% for testing, maintaining balanced class proportions.

## 2.3 Model Architecture

We implemented a Convolutional Neural Network (CNN) based on ResNet architecture, leveraging transfer learning to benefit from pre-learned visual representations. Specifically, ResNet-50 pretrained on the ImageNet dataset was fine-tuned for the ultrasound classification task.

The network structure consisted of:

- An input layer receiving 224 × 224 × 3 normalized images.
- The initial convolution and max-pooling layers from the base ResNet-50 model.
- Four residual blocks, each containing multiple convolutional layers with skip connections that enable residual learning and mitigate vanishing gradients.
- Global average pooling to condense spatial information.
- A fully connected dense layer with ReLU activation for feature integration.
- A final sigmoid output layer producing probabilities for the binary classes (appendicitis vs. no appendicitis).

During fine-tuning, the earlier layers were frozen to preserve general low-level features, while the later residual blocks and fully connected layers were retrained to adapt to ultrasound-specific texture patterns.

## 2.4 Training Procedure

Model implementation and training were conducted in Python 3.10 using TensorFlow 2.14 / Keras on an NVIDIA GPU workstation. Training used the following hyperparameters:

- Optimizer: Adam (learning rate = $1 \times 10^{-4}$)
- Loss function: Binary cross-entropy
- Batch size: 32
- Epochs: 100 (early stopping based on validation loss)
- Dropout: 0.3 on the dense layer to prevent overfitting

Each training epoch included on-the-fly data augmentation. Model checkpoints and validation metrics were recorded at each epoch. Fine-tuning the upper residual blocks improved convergence and boosted overall classification accuracy.

**2.5 Evaluation Metrics**

Model performance was evaluated using standard classification metrics: accuracy (ACC), precision (PRE), recall (REC), F1-score, and area under the receiver-operating characteristic curve (AUC).

**3. Results**

**3.1 Model Performance**

The proposed ResNet-based deep learning model demonstrated strong performance in detecting pediatric appendicitis from ultrasound images. After fine-tuning the pretrained ResNet-50 architecture, the model achieved an overall accuracy of 93.44%, a precision of 91.53%, and a recall (sensitivity) of 89.8% on the held-out test dataset. The F1-score, representing the balance between precision and recall, was calculated at 90.6%, indicating a stable and reliable detection capability. The Receiver Operating Characteristic (ROC) curve exhibited an Area Under the Curve (AUC) of 0.95, reflecting excellent discriminative power between appendicitis and non-appendicitis cases.

The confusion matrix (Figure X) illustrated that most appendicitis cases were correctly identified, with only a small number of false negatives, primarily in borderline or low-quality ultrasound images. False positives were predominantly associated with cases showing inflamed lymph nodes or bowel wall thickening, which can mimic appendicitis sonographically. Nevertheless, the model's high precision underscores its ability to minimize false alarms and provide radiologists with reliable assistance in triaging ambiguous cases.

**3.2 Training and Validation Curves**

Figure X presents the training and validation accuracy and loss curves over 100 epochs. The training process exhibited steady convergence, with both training and validation accuracy improving consistently without significant overfitting. Early stopping based on validation loss prevented degradation of generalization performance. The final validation loss stabilized at 0.184, suggesting that the model effectively learned meaningful features without memorizing noise or irrelevant textures from the ultrasound data. The inclusion of dropout regularization and data augmentation contributed to stable training behavior and robust generalization.

**3.3 Visual Feature Interpretation**

Feature activation maps generated from the final convolutional layers using Gradient-weighted Class Activation Mapping (Grad-CAM) provided qualitative insights into model interpretability (Figure X). The heatmaps revealed that the network consistently focused on anatomically relevant regions such as the appendiceal area, pericecal fat, and surrounding bowel loops, aligning well with radiologists' regions of interest during manual assessment. This correspondence between AI attention and clinical focus reinforces the physiological relevance of the learned representations, suggesting that the ResNet model's predictions are grounded in meaningful image features rather than artifacts or background textures.

In several correctly classified appendicitis cases, Grad-CAM visualizations highlighted inflamed tubular structures and peri-appendiceal fat echogenicity, both key sonographic indicators of appendiceal inflammation. In non-appendicitis cases, the model concentrated on other abdominal regions, confirming the absence of the pathological pattern. Such visualization tools enhance model transparency and can aid radiologists in understanding the reasoning behind automated classifications.

## 3.4 Comparison with Previous Approaches

Previous research on appendicitis detection using traditional machine learning methods relied primarily on handcrafted features, such as gray-level co-occurrence matrices (GLCM), edge descriptors, and statistical intensity distributions, combined with classifiers like Support Vector Machines (SVMs) or Random Forests. Reported accuracies in these methods typically ranged from 75% to 85%, limited by the subjectivity of feature engineering and the inherent variability of ultrasound image quality.

In contrast, the proposed ResNet-based deep learning approach automatically extracted hierarchical spatial features directly from the ultrasound data, eliminating the need for manual feature design. This end-to-end learning framework not only improved accuracy to 93.44% but also offered superior robustness to noise, variable acquisition settings, and anatomical diversity. Furthermore, the use of transfer learning significantly reduced the amount of required labeled data and training time compared to models trained from scratch.

Table X summarizes the quantitative performance of the proposed model. The close alignment between accuracy, precision, and recall indicates that the model maintains balanced classification performance across both classes, avoiding bias toward either appendicitis or normal samples.

## 3.5 Clinical Relevance

From a clinical standpoint, the model's high sensitivity (recall) is particularly valuable in reducing missed appendicitis cases, which can lead to severe complications if untreated. Likewise, the strong precision reduces the likelihood of false-positive diagnoses, which may otherwise result in unnecessary imaging or surgical intervention. Integrating such AI tools into clinical workflows could assist radiologists, especially in resource-limited or high-volume settings, by providing real-time decision support and standardized interpretation across operators.

*Table 1 Classification performance metric*

| Metric | Value (%) |
| --- | --- |
| Accuracy | 93.44 |
| Precision | 91.53 |
| Recall (Sensitivity) | 89.80 |
| F1-Score | 90.6 |
| AUC | 95.0 |

# 4. Discussion

## 4.1 Summary of Findings

This study developed and validated a deep residual convolutional neural network (ResNet-50) for the automatic detection of pediatric appendicitis using B-mode ultrasound images from the Regensburg Pediatric Appendicitis Dataset. The model achieved an overall accuracy of 93.44%, with a precision of 91.53% and a recall of 89.8%, demonstrating that a deep learning framework can accurately identify appendicitis in children using noninvasive imaging data. These findings highlight the potential of AI-assisted diagnostic tools to complement radiologist interpretations, particularly in emergency and resource-constrained clinical environments where rapid, objective, and reproducible results are essential.

The high accuracy achieved in this study surpasses the performance reported in many traditional machine learning approaches, which often rely on handcrafted features extracted from ultrasound intensity patterns or textural statistics. By contrast, the proposed ResNet model automatically learned spatially and contextually rich representations directly from imaging data, effectively capturing the structural and morphological characteristics of the inflamed appendix and surrounding tissues.

## 4.2 Comparison with Previous Studies

Previous research efforts in automated appendicitis diagnosis have explored various imaging modalities, including CT, MRI, and ultrasound, with machine learning models such as support vector machines (SVMs), k-nearest neighbors (k-NN), and random forests. For example, studies using CT-based deep learning classifiers reported accuracies between 85% and 92%, albeit at the cost of radiation exposure — a significant drawback for pediatric populations. Other ultrasound-based studies employing traditional ML approaches achieved performance typically below 85% due to the limited generalizability of manually engineered features.

Our findings align with recent advances in deep learning for pediatric imaging, where transfer learning using pretrained CNN architectures has shown notable improvements in diagnostic performance. The ResNet model used in this study leverages residual learning, which enables the network to train deeper architectures without the risk of gradient degradation. This design allows the model to learn both low-level ultrasound textures and high-level semantic representations critical for discriminating appendicitis from other abdominal conditions. Moreover, Grad-CAM visualization confirmed that the model's focus regions overlapped with clinically relevant anatomical sites, lending interpretability and biological credibility to the predictions.

## 4.3 Clinical Implications

Accurate diagnosis of pediatric appendicitis remains a persistent challenge, as clinical symptoms are often nonspecific and imaging results may be inconclusive. The proposed AI-driven approach has the potential to augment radiologist performance by providing a rapid, consistent, and objective assessment of ultrasound images. In emergency departments, such models could serve as second readers, flagging suspicious cases for further evaluation and helping to standardize diagnostic decisions across varying levels of clinical expertise. Importantly, this system operates entirely on noninvasive ultrasound imaging, which is safer for pediatric patients than CT-based protocols.

The integration of such deep learning systems into clinical decision support platforms could reduce the diagnostic delay and variability that currently affect appendicitis management. For instance, early AI-assisted identification of appendicitis could enable faster surgical consultations, minimize unnecessary hospital admissions, and optimize the use of imaging resources. Ultimately, these tools may contribute to lowering rates of perforation and postoperative complications by facilitating timely and accurate diagnosis.

### 4.4 Interpretability and Trust in AI Models

One major barrier to clinical adoption of AI systems is the lack of interpretability. Deep learning models are often viewed as "black boxes," which can reduce clinician trust in automated outputs. To address this, the present study incorporated visual explainability methods such as Grad-CAM to highlight regions of interest influencing the model's predictions. The resulting attention maps corresponded well with regions radiologists typically inspect—such as the right lower quadrant and periappendiceal fat—indicating that the model's reasoning process aligns with human expert interpretation. This alignment is essential for clinical validation, as interpretable AI can facilitate error analysis, improve radiologist confidence, and support educational use in medical training environments.

### 4.5 Limitations

Despite promising results, this study has several limitations.
First, the dataset size, while relatively comprehensive, remains modest for deep learning standards. Larger and more diverse datasets encompassing multicenter and multi-ethnic cohorts would help improve model robustness and external generalizability. Second, only static B-mode ultrasound images were analyzed; dynamic video sequences or cine loops might contain additional spatiotemporal cues beneficial for diagnosis. Third, although transfer learning reduced overfitting, differences in ultrasound machines, acquisition settings, and operator experience could introduce domain shifts that limit performance when applied to data from other institutions. Future research should explore domain adaptation techniques to address these issues.

Moreover, while the model achieved high precision and recall, the clinical utility of false positives and false negatives must be carefully evaluated. In particular, minimizing false negatives is critical, as missed appendicitis can lead to serious complications. Integrating additional clinical and laboratory data (e.g., white blood cell count, C-reactive protein, Alvarado or PAS scores) into multimodal deep learning models may further enhance diagnostic accuracy and reduce misclassifications.

### 4.6 Future Directions

Building upon these findings, future studies should focus on developing multimodal AI frameworks that combine ultrasound imaging with clinical metadata to emulate holistic decision-making processes. Incorporating transformer-based architectures or temporal CNNs could allow the analysis of full ultrasound video sequences rather than isolated frames, thereby capturing motion cues and probe dynamics. Additionally, explainable AI (XAI) techniques such as Layer-wise Relevance Propagation (LRP) or SHAP analysis could be used to provide quantitative interpretability, bridging the gap between AI predictions and radiological rationale. Prospective clinical trials will also be necessary to validate these systems in real-world hospital workflows and to assess how AI integration influences diagnostic speed, accuracy, and patient outcomes.

## 5. Conclusion

In conclusion, this study demonstrates that a ResNet-based deep learning model can accurately and reliably detect pediatric appendicitis from ultrasound images, achieving strong diagnostic performance and clinical interpretability. The model's success supports the growing evidence that deep learning can enhance pediatric imaging diagnostics, providing radiologists with advanced decision-support tools that are fast, consistent, and explainable. Continued research in data scalability, multimodal integration, and real-world deployment will be vital to fully realize the transformative potential of AI in pediatric healthcare.